\documentclass[apjl]{emulateapj}

\shorttitle{A genuine intermediate-Age Globular Cluster in M33}

\shortauthors{Chandar et al.}

\begin{document}
\title{A genuine intermediate-Age Globular Cluster in M33\altaffilmark{*}}

\author{Rupali Chandar\altaffilmark{1}, Thomas H. Puzia\altaffilmark{2}, Ata
  Sarajedini\altaffilmark{3}, \& Paul Goudfrooij\altaffilmark{2}} 
\altaffiltext{*}{Based on observations obtained with the Hectospec instrument
  at the MMT Observatory. The MMT Observatory is a joint venture of the
  Smithsonian Institute and the University of Arizona. Based on observations
  made with the NASA/ESA Hubble Space Telescope, obtained from the data
  archive at the Space Telescope Science Institute. STScI is operated by the
  Association of Universities for Research in Astronomy, Inc. under NASA
  contract NAS 5-26555.} 
\altaffiltext{1}{Department of Physics and Astronomy, Johns Hopkins
  University, 3400 North Charles Street, Baltimore, MD 21218, USA. 
  {\tt rupali@pha.jhu.edu}} 
\altaffiltext{2}{Space Telescope Science Institute, 3700 San Martin Drive,
    Baltimore, MD 21218, USA. {\tt tpuzia@stsci.edu, goudfroo@stsci.edu}}
\altaffiltext{3}{Department of Astronomy, University of Florida, 211 Bryant
  Space Science Center, P.O. Box 112055, Gainesville, FL 32611, USA. 
  {\tt ata@astro.ufl.edu}}

\begin{abstract}
We present deep integrated-light spectroscopy of nine M33 globular
clusters taken with the Hectospec instrument at the MMT Observatory.
Based on our spectroscopy and previous deep color-magnitude diagrams
obtained with HST/WFPC2, we present evidence for the presence of a genuine
intermediate-age globular cluster in M33. The analysis of Lick line
indices indicates that all globular clusters are metal-poor
([Z/H]~$\la-1.0$) and that cluster M33-C38 is $\sim\!5\!-\!8$ Gyr younger
than the rest of the sample M33 star clusters. We find no evidence for a
population of blue horizontal branch stars in the CMD of M33-C38, which
rules out the possibility of an artificially young spectroscopic age due
to the presence of hot stars. We infer a total mass of
$5-9\times10^{4}~M_{\odot}$ for M33-C38, which implies that M33-C38 has
survived $\sim2-3$ times longer than some dynamical evolution model
predictions for star clusters in M33, although it is not yet clear to
which dynamical component of M33 -- thin disk, thick disk, halo -- the
cluster is associated.
\end{abstract}


\section{Introduction}

Star clusters which have survived for several ($\sim2-8$) billion years
(intermediate-age clusters) can provide important clues concerning both
the evolution history of their parent galaxies and for understanding the
destruction processes which erode star cluster systems \citep[e.g.,][and
references therein]{goudfrooij04}. In the Galaxy, clusters of this age all
fall in the category of ``old open clusters'' \citep[see compilation
in][]{friel95} -- they all reside in the Galactic thin disk, and have
lower masses ($\sim$few $10^3$~$M_{\odot}$) than typical ancient globular
clusters ($\ga10^{4.5}~M_{\odot}$). The oldest of these old open clusters
help to pinpoint the age of the thin disk, while the ages and masses of
the entire population help constrain the survival rates of clusters in the
Galactic disk. Intermediate-age clusters are also known to exist in the
Large and Small Magellanic Clouds (LMC and SMC). The LMC contains a
population of clusters which are $1-3$~Gyr, while the SMC formed numerous
clusters $4-8$~Gyr ago.

In more distant galaxies, it becomes increasingly difficult to establish
the presence of intermediate-age clusters, primarily because the
techniques available to study compact clusters become more limited. For
example, the integrated optical colors of intermediate-age clusters are
degenerate with those of ancient ($\geq8$~Gyr) clusters. Although a number
of clusters have absorption line index strengths (measured from
integrated-light spectroscopy) which suggest that these objects are of
intermediate age, their absolute ages cannot be definitively established
due to the possible presence of hot stars in the core helium burning stage
(hot ``horizontal branch'' or HB stars). These hot HB stars could
potentially boost the Balmer absorption line strengths sufficiently to
mimic younger ages for old globular clusters \citep[e.g.,][]{lee05,
maraston00}. Ideally, cluster ages should be determined from deep
color-magnitude diagrams (CMDs) which reveal the main sequence turnoff; to
date however, this has only been accomplished for a single cluster beyond
the Magellanic Clouds \citep[M31-SKHB 312;][]{brown04}, using 129 orbits
of {\it Hubble Space Telescope} (HST) time.

In this Letter, we present integrated-light (optical) spectroscopy for
nine ancient star cluster candidates in the nearby spiral galaxy M33.
These clusters have deep CMDs available from HST/WFPC2 observations, which
clearly reveal the HB morphology. The combination of HB morphology and
absorption line measurements result in robust relative age estimates, and
are ideal to search for intermediate-age clusters in M33.


\section{Data and Analysis}

\subsection{Integrated-Light Spectroscopy}

We obtained integrated spectroscopy for $\sim\!150$ star clusters and
cluster candidates in M33 using the MMT/Hectospec instrument
\citep{fabricant05} at a dispersion of 1.2~\AA~pix$^{-1}$. In addition to
clusters, a significant number of background regions were chosen, in order
to sample the range of background light provided by the galaxy itself. The
clusters were observed on the night of October 27/28, 2005, with four 1100
second exposures (for a total of 73.3 minutes). Each of the 300 Hectospec
fibers subtends $1.5\arcsec$ on the sky. The data were reduced using the
HSRED reduction pipeline developed by R. Cool\footnote{Available from
http://mizar.as.arizona.edu/\~{}rcool/hsred.}. All observations were bias
subtracted, overscan corrected, and trimmed. The science exposures were
flat-fielded and combined together to eliminate cosmic rays, and one
dimensional spectra were extracted and wavelength calibrated. The final
resolution measured from the FWHM of HeNeAr comparison spectra was
$\sim4.7$~\AA. Spectra for the entire M33 cluster sample will be presented
in a future work. Here, we focus on a subset of nine clusters which have
available CMDs reaching below the HB \citep[hereafter S00]{sarajedini98, sarajedini00}.

\begin{figure}
\centering
\includegraphics[bb= 0 0 400 390, width=8cm]{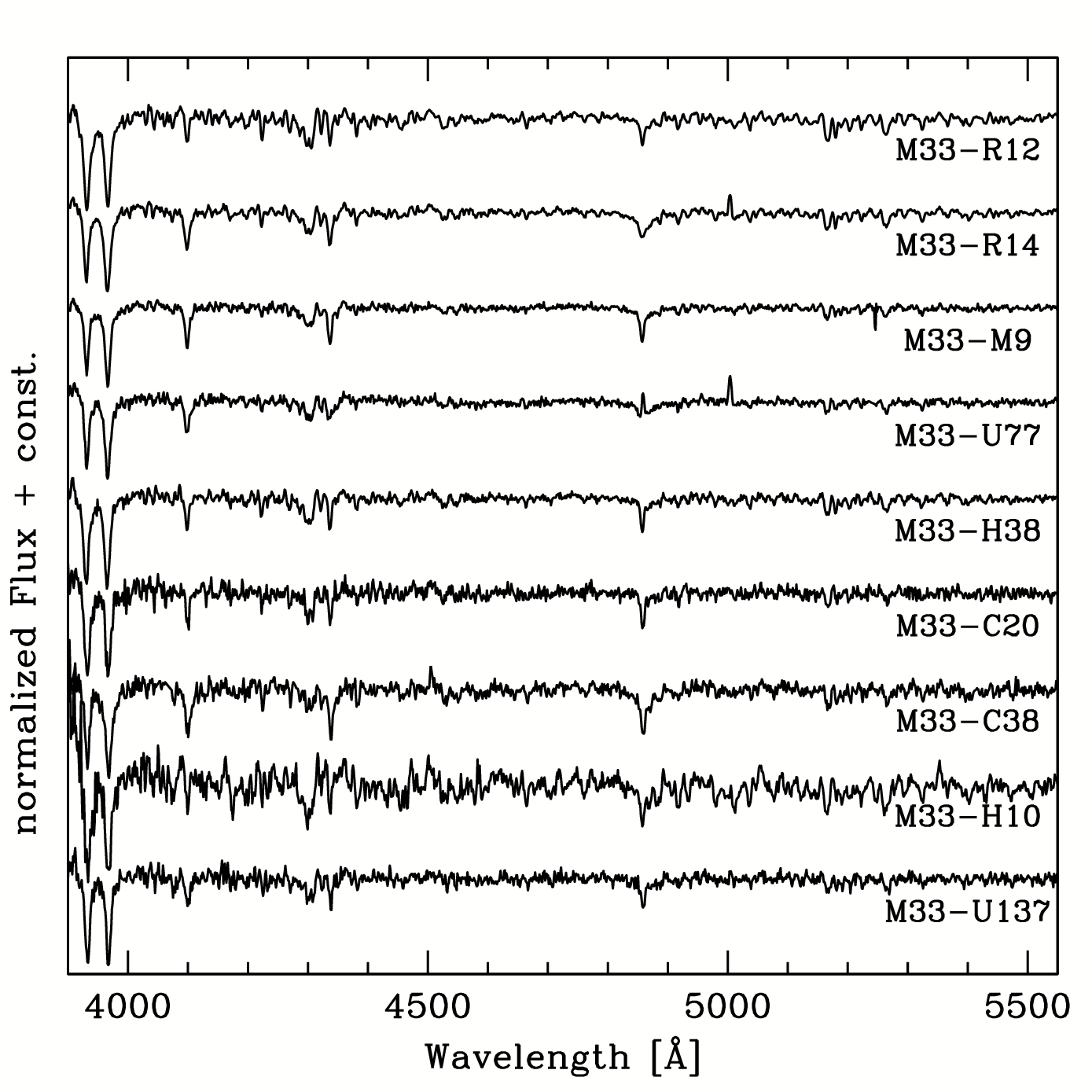}
\caption{The blue portion of the background subtracted spectra of nine 
M33 globular clusters, taken with MMT/Hectospec are shown. The spectra
have been normalized by a spline fit to their continua.}
\label{fig1}
\end{figure}

Because the underlying galaxy light within each object fiber can affect
the absorption line strengths, we created a mean background spectrum for
each cluster.  This was accomplished by averaging several background
spectra chosen to match the galaxy background level measured in an annulus
around each cluster from reduced M33 images provided by the Local Group
Survey\footnote{http://www.lowell.edu/\~{}masey/lgsurvey.}). These
customized background spectra were then subtracted from each object.  For
the nine clusters presented here, the cluster spectra have such high
signal-to-noise, that the background subtraction has little impact. In
Figure~\ref{fig1} we present integrated, background-subtracted spectra of
the globular clusters discussed in this work.

The relative ages and abundances of stellar populations can be estimated
from integrated-light spectra, by comparing the absorption line strengths
for age and metallicity-sensitive lines. The measurements and accompanying
uncertainties are computed using the techniques described in
\cite{puzia02}, for line indices defined in \cite{worthey94} and
\cite{worthey97}. The passband definitions were shifted to account for the
radial velocity of each cluster, assuming the measurements given in
\cite{chandar02}. Because no observations of Lick standard stars have been
made to date with MMT/Hectospec, we do not calibrate the index
measurements to the Lick/IDS system \citep{burstein84}, and instead focus
on {\it relative} cluster ages. A comparison of the line indices measured
for the nine clusters presented in this work with predictions of the
\cite{thomas03, thomas04} models is illustrated in Figure~\ref{ps:am}.

\subsection{Color-Magnitude Diagrams}
The CMDs for the nine clusters analyzed in this work were originally
presented in Sarajedini et al. (2000; hereafter S00). These are based on
$V_{\rm F555W}$ and $I_{\rm F814W}$ observations taken with the HST/WFPC2
instrument. We refer the reader to S00 for details of data acquisition,
reduction, and analysis. The CMDs show that only M9 and U77 have blue HB
stars; the rest of the M33 clusters in the S00 sample have HBs which lie
completely redward of the RR Lyrae instability strip, in the so-called
``red clump''.

Here, we present new CMDs based on the S00 data for two clusters of
particular interest -- M9 and C38. These have been radially cleaned, so
that blends and other contaminants from the crowded central regions are
not included in the CMD. In Figure~\ref{fig:cmd} we show the new CMDs for
these two clusters which are further discussed below.

\section{Discussion}

\subsection{Evidence for an Intermediate-Age Cluster}
\begin{figure*}
\centering
\includegraphics[bb= 0 0 400 400, width=5.5cm]{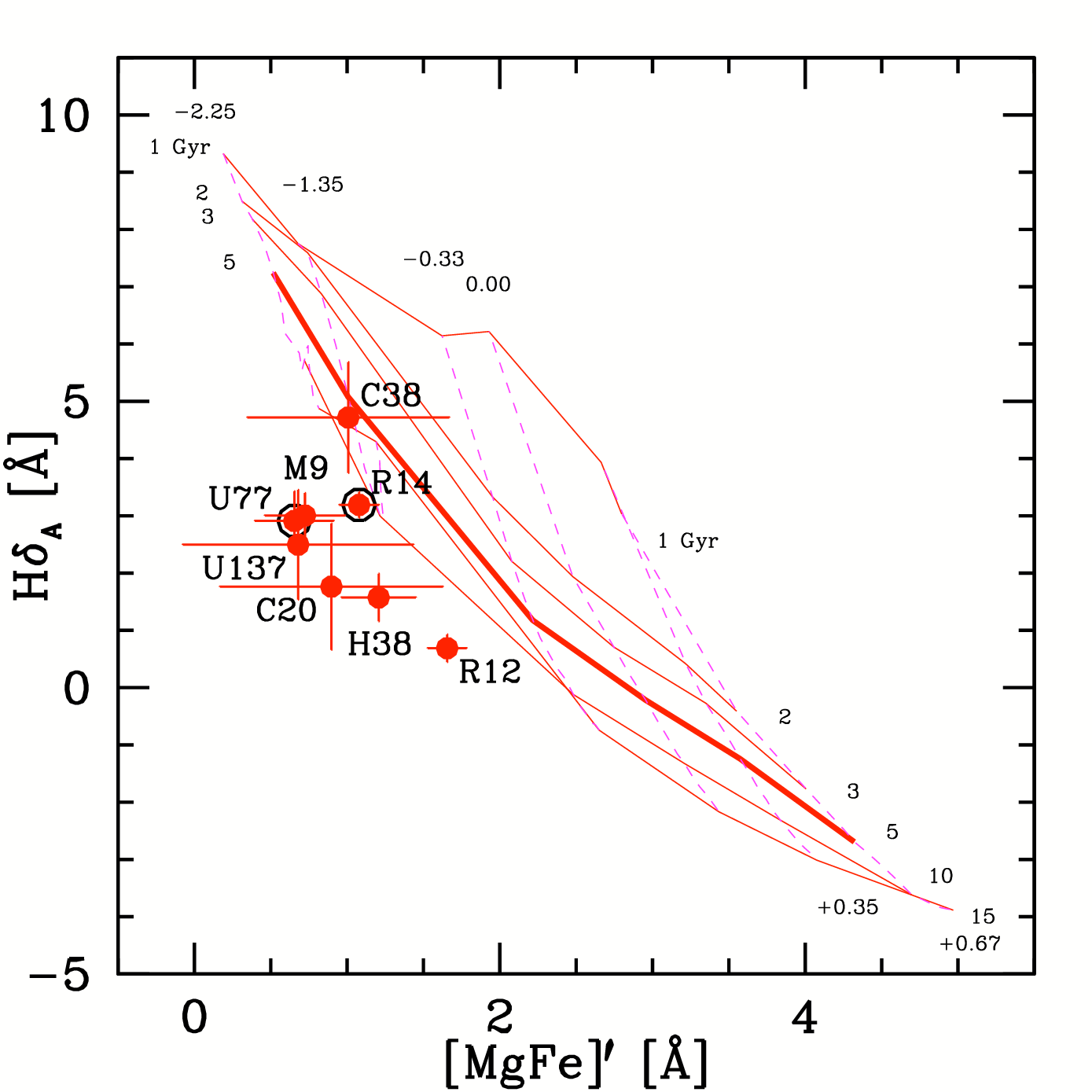}
\includegraphics[bb= 0 0 400 400, width=5.5cm]{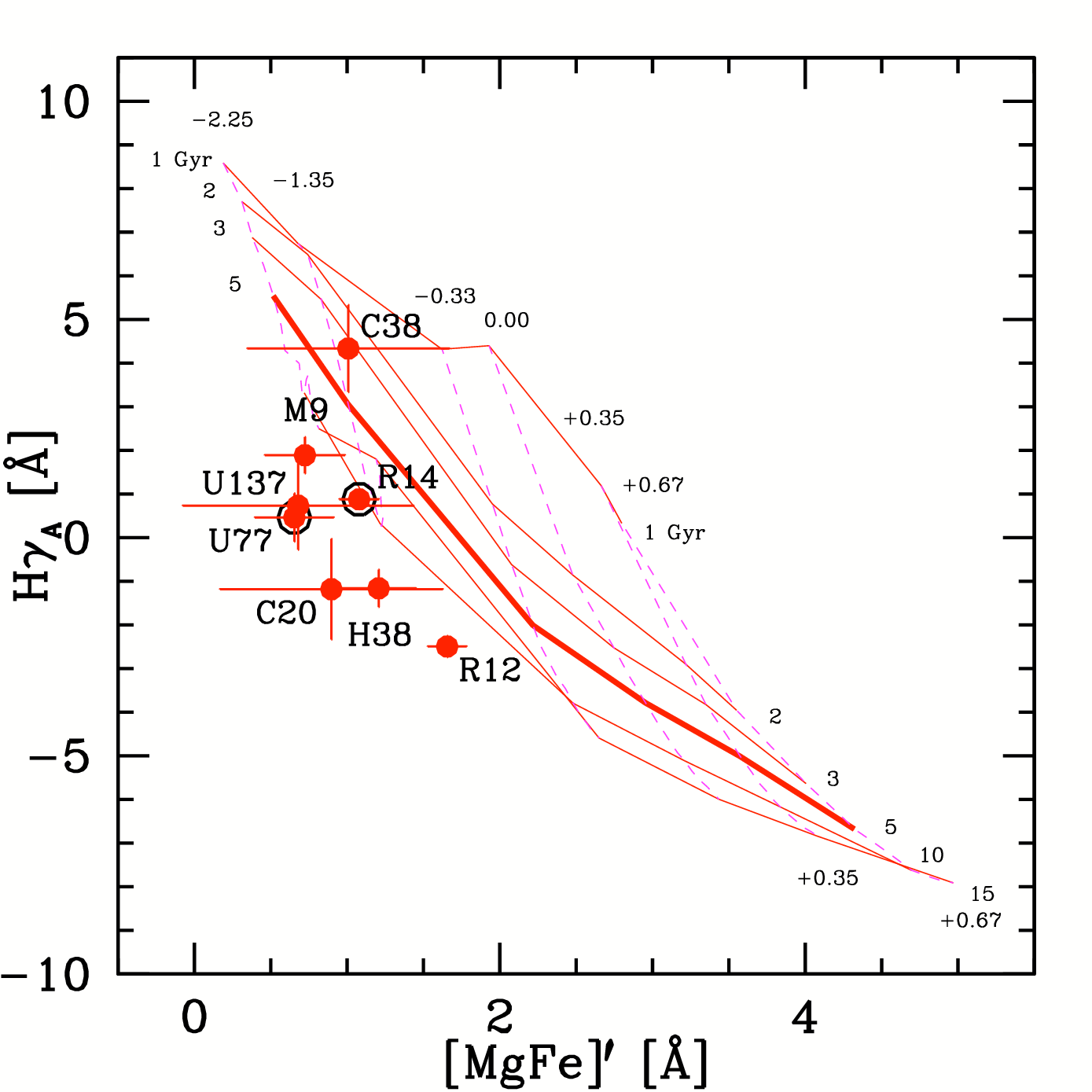}
\includegraphics[bb= 0 0 400 400, width=5.5cm]{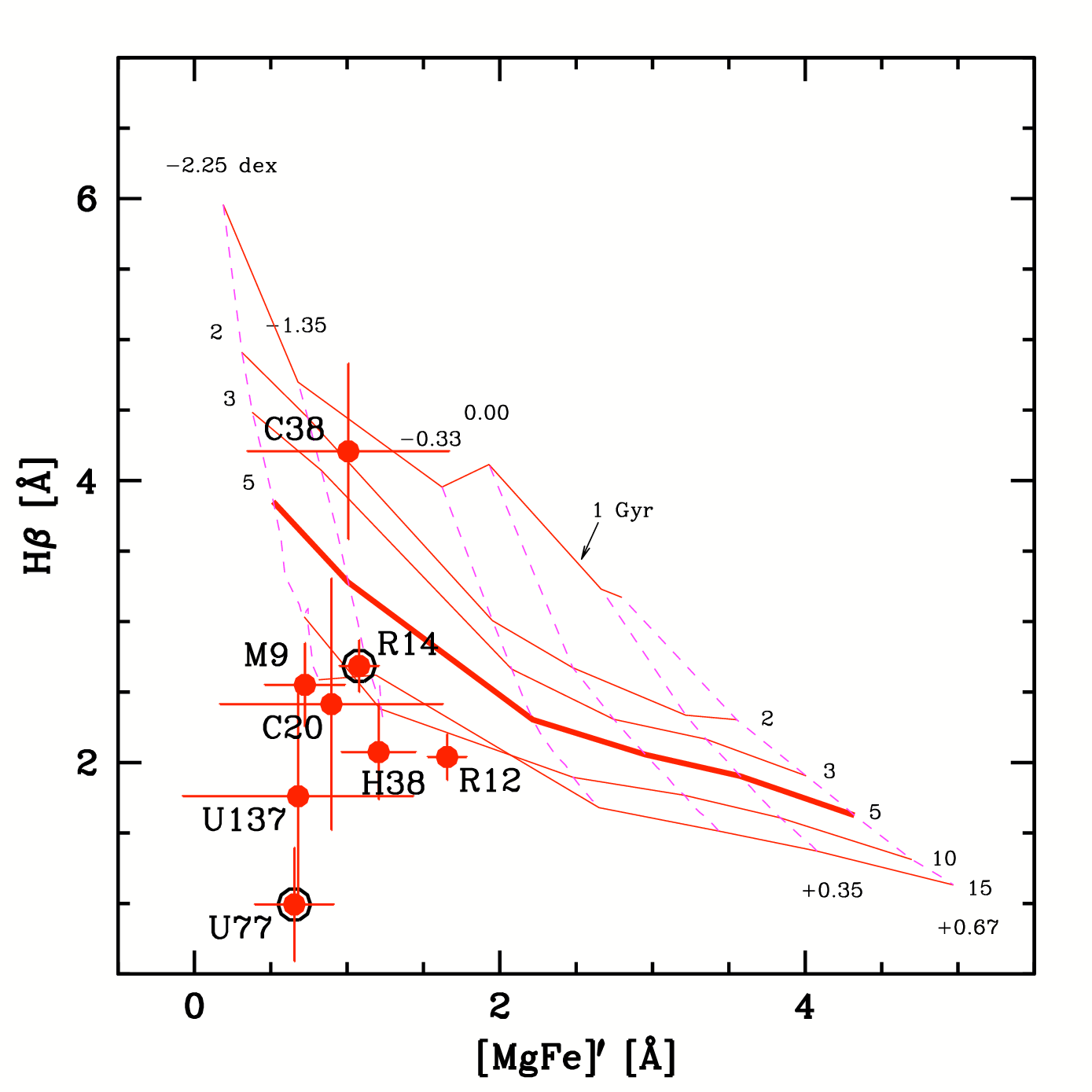}
\caption{Age-metallicity diagnostic plots for the sample M33 globular
clusters, constructed from three different Balmer line indices:
H$\delta_{\rm A}$ ({\it left panel}), H$\gamma_{\rm A}$ ({\it middle
panel}), and H$\beta$ ({\it right panel}) vs.~[MgFe]\arcmin\
\citep[see also Fig.~9 of][for details]{puzia05}. Over-plotted are SSP models of
\cite{thomas04} for [$\alpha$/Fe]~$=0.3$, metallicities [Z/H]=$-2.25,
-1.35, -0.33, 0.00$, 0.35, and 0.67 dex ({\it dashed lines}), and ages
15, 10, 5, 3, 2, 1 ({\it solid lines}). The thick solid line is the 5
Gyr isochrone, and is used to split between old and intermediate age
globular clusters. Note that some clusters are not plotted in the
individual panels due to contamination of their index passband
definitions by bad pixels. Clusters with H$\alpha$ observed in
emission are indicated by open circles. All clusters are labeled as in
Fig.~\ref{fig1}. We point out that the Lick index measurements are not
calibrated to the Lick system so that only relative measures should be
derived from the shown diagrams.}
\label{ps:am}
\end{figure*}

In Figure~\ref{ps:am}, we show the measurements for the metal-sensitive
absorption line index [MgFe]\arcmin\ versus the age-sensitive Balmer line
indices H$\delta_{\rm A}$, H$\gamma_{\rm A}$, and H$\beta$, as is
typically done to estimate the ages and chemical compositions of stellar
populations from integrated-light spectroscopy \citep[e.g.,][]{puzia05}.
For comparison, we show simple stellar population (SSP) model predictions
from \cite{thomas04} for six different ages (solid lines) and six
different values of [Z/H] (dashed lines). We also compared the M33 cluster
measurements with the solar-scaled SSP model predictions of \cite{bc03} at
4.7~\AA\ resolution (not shown), which yield results consistent with the
\cite{thomas04} models.~All panels in Figure~\ref{ps:am} show a ``cloud''
of points located towards low metallicities and old ages, although we
cannot rule out an age spread of the order of $\Delta t/t\la 0.3$, which
translates into a $\sim\!4$~Gyr spread at 12 Gyr absolute age. Cluster C38
is clearly offset from the ``cloud" in every panel, and clearly appears to
have a younger age than the other clusters -- recall that our measurements
have not been absolutely calibrated to the models, but should be robust in
a relative sense. Therefore, Figure~\ref{ps:am} implies that C38 is
$\sim\!5-8$ Gyr younger than the other M33 clusters plotted here. Clusters
with H$\alpha$ seen in emission (R14 and U77) are plotted with different
symbols in Figure~\ref{ps:am}. The Balmer absorption-line strengths for
these clusters thus represent lower limits.

\begin{figure}[!t]
\centering
\includegraphics[bb=46 291 585 752,width=8.5cm]{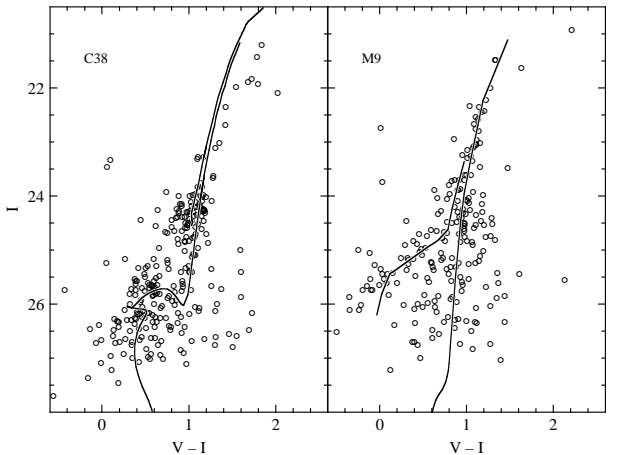}
\caption{$V-I$ versus $I$ CMD for M33 clusters C38 and M9. The
CMDs have been restricted to include stars in an annulus between 1 and
2\arcsec\ away from the center of each cluster, in order to minimize
blending and contamination. A Girardi isochrone for an age of 2 Gyr and
$Z=0.004$ is overplotted onto the C38 CMD, which shows the red clump stars
and a hint of a main sequence turnoff. A fiducial HB for the Galactic
globular cluster M3 \citep{johnson98} is overplotted onto the M33-M9 CMD,
which clearly reveals the blue horizontal branch stars in this cluser.}
\label{fig:cmd}
\end{figure}

Could the stronger Balmer line indices for C38, which suggest a younger
age, be due to the presence of blue/hot HB stars? In Figure~\ref{fig:cmd},
we show the resolved CMD of C38 based on deep HST/WFPC2 observations for
this cluster, and also for M9 (one of the older clusters) for comparison.
C38 shows no blue HB stars, while M9 does. Note that the HST observations
have sufficient depth and resolution to detect blue HBs if they are
present, but they clearly do not exist in C38. Therefore, the presence of
hot HB stars, which has been suggested as a plausible explanation for
apparent intermediate ages of clusters in different galaxies \citep[such
as in M31;][]{puzia05b}, cannot be the correct explanation in this case.
The line indices and the knowledge of the HB morphology taken together,
indicate that it is the {\it age\/} of the cluster which drives the
observed Balmer line strengths.

There are two features from the resolved CMD of C38 shown in
Figure~\ref{fig:cmd} which add weight to this conclusion.  First, S00
found that the absolute I band magnitude of the red clump for this cluster
is more luminous than all of the others in their sample, and that this
behavior is reasonably well fit by the $\sim2$~Gyr theoretical HB models
of \cite[see also Figure~26 in S00]{girardi99}.
Secondly, when the CMD for C38 is cleaned to exclude the most crowded
central regions, there appears to be a hint of a main sequence turnoff
(MSTO) in the CMD, as illustrated in the left panel of
Figure~\ref{fig:cmd}. In contrast, M9 does not show evidence for a
MSTO.~Also in this figure, we show a Z = 0.004 isochrone at an age of
2~Gyr from Girardi et al. (2000) for comparison. Taken together with the
spectroscopic results and the abnormally bright red clump, the presence of
a MSTO in the CMD of C38 provides strong evidence that C38 is a genuine
intermediate-age ($\sim2-5$~Gyr) star cluster in M33.

\subsection{Properties of M33-C38}

Here we compile known properties of M33-C38. This cluster lies at a
projected distance of $\sim\!4.5$ kpc from the center of M33, has a total
$V$-band magnitude of 18.01 and $B-V=$ 0.73 and $V-I=0.89$ colors
\citep{christian88, chandar99, chandar02}. S00 estimated
that it has an [Fe/H] of $-0.65\pm0.16$ from the slope of the red giant
branch, which is consistent with our spectroscopy given the uncertainties
([Fe/H]$=-1.1\pm0.6$).

\cite{chandar02} measure a radial velocity of $-145\pm30$ km/s for this
cluster. The local H\,{\sc i} (disk) velocity at the location of C38 is
found to be $\sim-100$~km/s from the radio maps of \cite{warner73}.
Although the cluster velocity measurement deviates from the local disk
motion in Figure~\ref{fig:cmd} of \cite{chandar02}, (showing a comparison
of cluster velocities versus the local disk motions relative to cluster
age, C38 falls in a region which is ambiguous as to which M33 component
C38 belongs, clusters with the most deviant velocities are almost certainly
in a halo/thick disk component). Therefore, this object could belong to
either the thin disk or a halo/thick disk component. Implications for
these possibilities are discussed in the next section.

Unfortunately M33-C38 does not have any published $M/L$ ratios, which
would provide a direct estimate of its total mass. There are, however,
published $M/L_V$ values for three of the clusters presented in this work
\citep[H38, M9, and R12;][]{larsen02}. They have an average
$M/L_V=1.53\pm0.18$, which is consistent with $M/L_V$ measurements for
Galactic GCs \citep{mclaughlin00}. The integrated spectra presented here
corroborate that these objects are ancient, even though their exact ages
are not known. Given the fact that $M/L_V$ {\it decreases} with decreasing
age, as an upper limit for C38 we take the highest value found empirically
by \cite{larsen02} of $M/L_V\!=\!1.87$. As a lower limit, we assume the
$M/L_V$ value predicted by the \cite{bc03} models at an age of 2~Gyr and
for a low metallicity system ($M/L_V\!=\!1.125$). Assuming a foreground
reddening $E_{B-V}=0.1$ mag towards M33 and a distance modulus of 24.64
mag, we estimate that C38 has a mass in the range $5-9\times10^{4}\,
M_{\odot}$. This is about an order of magnitude higher than the masses for
old Galactic open clusters, and within the range of ancient globular
clusters.

\subsection{Implications of Intermediate-Age Clusters in M33}

The confirmed presence of an intermediate-age cluster in M33 has important
implications for star/cluster formation processes in spiral galaxies,
regardless of which structural component C38 belongs to. Below we explore
the implications of the presence of intermediate-age clusters in both a
thick disk versus halo component, and also in the thin disk.

If C38 belongs to either a thick disk or halo component, it is likely that
at least a few of the other clusters in our sample belong to the same
structural component, so that a relatively large age spread among M33
clusters would be present in that component. In the Galaxy, halo GCs are
known to be old and have an age spread of only $\sim\!3$~Gyr, whereas the
GCs associated with the thick disk are essentially coeval
\citep{deangeli05}. We suggest that a large age spread for M33 clusters
would favor a halo over a thick disk origin, for the following reasons.
The thick disk in the Milky Way is relatively metal-rich, and believed to
have formed during a single event, likely the accretion of a relatively
massive satellite galaxy which puffed up the disk
\citep[e.g.][]{morrison93, abadi03, martin04}. Such an origin for a thick
disk in M33 would imply that clusters in M33 formed for many Gyr
relatively undisturbed in the disk, until several Gyr ago when an
accretion event took place, resulting in a thick disk with a cluster
population spanning a large range of ages. Given that to date there is no
evidence for such an event in M33, and that this type of scenario would
require that no significant merging occurred before or since, we suggest
that it is unlikely that clusters residing in a thick disk would have an
age spread of several Gyr.

Therefore, the presence of C38 seems to favor a halo component over a
thick disk component. This would point to a longer timescale for the
buildup of the M33 halo relative to that found for the Milky Way, as
suggested previously by S00 and \cite{chandar02}.

If instead of the halo/thick disk, C38 resides in the thin disk of M33 (the
velocity measurements are ambiguous in discriminating between these two
possibilities), the intermediate age of this object also has important
implications. First, this would confirm the presence of relatively
massive, intermediate-age disk clusters, which are not widely observed in
our own Galaxy (but might be hidden behind large columns of dust). Second,
the ability of massive clusters $(\sim\!10^4~M_{\odot}$) to survive in the
disk of M33 has been suggested to be limited to $<1$~Gyr. \cite{lamers05}
used cluster samples in M33 and their theory of cluster dissolution
\citep[e.g.][]{boutloukos03} to estimate directly from the cluster age and
mass distributions the ability of clusters to survive in different
environments. For M33, they find that $10^4~M_{\odot}$ clusters will
disrupt in $\log({\rm age})=8.8$ years ($\sim\!600$~Myr), and that a
$5\times10^{4}\, M_{\odot}$ cluster will disrupt in $\sim1.5$~Gyr.~Mass
estimates for C38 put it at $5-9\times10^{4}\, M_\odot$. Although we do
not know its precise age, it appears that C38 has likely survived $2-3$
times longer than this prediction. Regardless, if C38 is part of M33's
thin disk, this points to the formation of a star cluster which is more
massive than known counterparts of a similar age in the Local Group, and
also to the ability of such an object to survive disruption forces in the
environment of M33 for several billion years.

\vspace{-0.5cm}

\acknowledgments  We thank D. Fabricant and N. Caldwell
for making the wonderful Hectospec instrument available to the community.
R.C. and T.H.P. acknowledge support by NASA through grants GO-10402-A
and GO-10129 from the Space Telescope Science Institute, which is operated
by AURA, Inc.,~under NASA Contract NAS5-26555. A.S. gratefully
acknowledges support from NSF CAREER grant AST-0094048.

\end{document}